\documentclass[%
 aip,
 amsmath,amssymb,
reprint, onecolumn 
]{revtex4-1}
\numberwithin{equation}{section}
\renewcommand{\theequation}
{\arabic{section}.\arabic{equation}}
\newcommand{\sech}{\text{sech}}
\usepackage{graphicx}
\usepackage{dcolumn}
\usepackage{bm}

\usepackage[utf8]{inputenc}
\usepackage[T1]{fontenc}
\usepackage{mathptmx}
\usepackage{etoolbox}

\makeatletter
\def\@email#1#2{%
 \endgroup
 \patchcmd{\titleblock@produce}
  {\frontmatter@RRAPformat}
  {\frontmatter@RRAPformat{\produce@RRAP{*#1\href{mailto:#2}{#2}}}\frontmatter@RRAPformat}
  {}{}
}%
\makeatother
\begin{document}

\preprint{AIP/123-QED}

\title[]{Painlev\'e Analysis, Prelle-Singer Approach, Symmetries \\and Integrability of Damped H\'enon-Heiles System}

\author{C. Uma Maheswari}%
 \affiliation{Ramanujan Institute for Advanced Study in Mathematics, University of Madras, Chennai-600005, Tamil Nadu, India}
\email{ramajayamsaha@gmail.com}
 
\author{N. Muthuchamy}%
 \affiliation{Ramanujan Institute for Advanced Study in Mathematics, University of Madras, Chennai-600005, Tamil Nadu, India}

\author{V. K. Chandrasekar}
 
\affiliation{%
Centre for Nonlinear Science and Engineering, School of Electrical and Electronics Engineering, SASTRA Deemed University, Tanjavur-613401, Tamil Nadu, India
}%

\author{R. Sahadevan}%
 \affiliation{Ramanujan Institute for Advanced Study in Mathematics, University of Madras, Chennai-600005, Tamil Nadu, India}

\author{M. Lakshmanan}
 
\affiliation{%
Centre for Nonlinear Dynamics, School of Physics, Bharathidasan University, Tiruchirappalli-620024, Tamil Nadu, India
}%

\begin{abstract}
	\textbf{ABSTRACT}	\\We consider a modified damped version of H\'enon-Heiles system and investigate its integrability. By extending the Painlev\'e analysis of ordinary differential equations we find that the modified H\'enon-Heiles system possesses the Painlev\'e property for three distinct parametric restrictions. For each of the identified cases, we construct two independent integrals of motion using the well known Prelle-Singer method. We then derive a set of nontrivial non-point symmetries for each of the identified integrable cases of the modified H\'enon-Heiles system. We infer that the modified H\'enon-Heiles system is integrable for three distinct parametric restrictions. Exact solutions are  given explicitly for two integrable cases.
\end{abstract}

\maketitle
	
\section{Introduction}\label{sec1}

	 The study of nonlinear differential equations has attracted much attention due to various reasons during the past several decades\cite{Abl1,Lak,Inc}. Nonlinear differential equations are not solvable or integrable in general and they possess various exciting mathematical structures and physical features. Given a nonlinear differential equation, there exists no unique method to determine whether  it is integrable or not. There exist several definitions for integrability such as Lax integrability, algebraic integrability, Liouville integrability, etc. in the literature\cite{Lak}.  Several analytic and numerical methods have been devised by different research groups to deal with nonlinear differential equations in general and integrability in particular\cite{Abl1, Abl2, Lak, Saha1, Uma}. Among them, the Painlev\'e analysis proposed by Ablowitz, Ramani, and Segur for ordinary differential equations (ODE) and extended to partial differential equations (PDE) by Weiss, Tabor and Carnevale played a vital role in identifying the integrable cases of  both nonlinear ODEs and PDEs during the past few decades \cite{Abl, Gra, Wei, Inc, Bountis}.  The Painlev\'e test is based on the realization that the integrability of differential equations is known to be related to  the singularity structure of the solution. A nonlinear differential equation is said to possess the Painlev\'e property if its general solution is single valued about the movable singular point in the case of ODE and movable singularity manifold for PDEs. A nonlinear dynamical system governed either by ODE or PDE passes the Painlev\'e test then it is expected  to be integrable and have a solution that can be described by an appropriate Laurent series expansion locally \cite{Clar, Inc, Laksh}. The effectiveness of the Painlev\'e analysis has been widely demonstrated by  identifying several new integrable systems \cite{Laksh,Ray,Pet}. Another important problem of nonlinear differential equations is to find its integrals of motion. If an $N^{th}$ order nonlinear ODE admits $N-1$ functionally independent integrals then it is completely integrable. Also, the existence of integrals of motion enables one to reduce the order of the differential equation. Several analytic methods have been devised to find integrals of motion for nonlinear differential equations \cite{Lak,Pre,Pol,Por,Mah}. Darboux has proposed a systematic method to construct integral of motion for Hamiltonian systems with two degrees of freedom. Chandrasekhar has successfully exploited the idea of Darboux to construct second and third integrals of motion for Hamiltonian systems with three degrees of freedom \cite{Cha, Whi}. Recent investigations reveal that the direct method proposed by Prelle and Singer provides an efficient tool to find integrals of motion for coupled nonlinear second order ODEs \cite{Vkc1, Vkc2, Pre}. Also, it is known that there exists a deep connection between the existence of nontrivial symmetries and integrals of motion of nonlinear differential equations\cite{ Saha,Blu}.\\	
	\noindent It is known that the celebrated H\'enon-Heiles (HH) system described by the Hamiltonian 
	
	\begin{equation}
		H = \frac{1}{2}\Bigg(\left(\frac{dx}{dt}\right)^2+ \left(\frac{dy}{dt}\right)^2\Bigg) + \frac{1}{2}\Big(Ax^2+By^2\Big)+\Big(\alpha x^2-\frac{1}{3}\beta y^2\Big) y,
	\end{equation}
	with equations of motion	
	\begin{equation}
		\frac{d^2 x}{dt^2}	 + A x + 2 \alpha x y  =0,
	\end{equation}
	\begin{equation}
		\frac{d^2 y}{dt^2}	 + B y + \alpha x^2 - \beta y^2 =0,	
	\end{equation}	
	where $A,B,\alpha$ and $\beta$ are constants is completely integrable in the sense of Liouville for the following three distinct cases:\\
	$(i) \ A=B, \ \ \ \beta = -\alpha$. \\
	$ (ii) \ A,B \ arbitrary, \ \ \ \beta = -6\alpha$.\\
	$(iii) \ B = 16 A, \ \ \ \beta = -16\alpha. $\\
	If an integrable nonlinear dynamical system undergoes perturbation then the resulting system may not be integrable in general. With this aim, we introduce a damping term in the equations of motion of HH system in the following form:
	\begin{equation} \label{a}
		\frac{d^2 x}{dt^2} +  \phi_1 \frac{dx}{dt} + A x + 2 \alpha x y  =0,
	\end{equation}
	\begin{equation}\label{b}
		\frac{d^2 y}{dt^2} +  \phi_2 \frac{dy}{dt} + B y + \alpha x^2 - \beta y^2 =0,
	\end{equation}
	where $ \phi_1$ and $ \phi_2$ are constants in addition to with $A,B,\alpha$ and $\beta.$ We refer to the above as the modified H\'enon-Heiles system equations. We also note here that one can always use the scaling transformation available for the independent variable t to fix the parameter $\alpha=1$ without loss of generality and also to agree with the original H\'enon-Heiles system in the absence of damping. However, we prefer to keep $'\alpha'$ as arbitrary here for convenience in order to compare with the various works in the literature on the generalization of H\'enon-Heiles system and one can always fix $\alpha=1$ whenever necessary.  \\
	\indent  The aim of this article is to identify the parametric restrictions at which (i) the general solution of it is single valued about the movable singular point  (ii) admits two functionally independent integrals of motion and (iii) admits more than one set of nontrivial non-point symmetries through the Painlev\'e analysis, Prelle-Singer method and Lie symmetry analysis, respectively.  The plan of the article is as follows. In section $2,$ we extend the Painlev\'e analysis of ODEs to the modified H\'enon-Heiles system governed by the above system  $\big($equations (\ref{a}) and (\ref{b})$\big)$ of two-coupled second order ODEs and show that its solution is single valued about the movable singular point and admit four arbitrary constants for three distinct parametric restrictions. In section $3,$ we apply the Prelle-Singer method to the system of two coupled second order ODEs associated with the modified H\'enon-Heiles system and derive two independent integrals for each of the identified cases ensuring their integrability. In section $4,$ we derive a set of nontrivial non-point symmetries for each of the identified integrable cases of the modified H\'enon-Heiles system. An attempt is made to find exact solution, if it exists, for each one of the three identified cases in section $5.$  A summary of the results of our investigation is given in section $6.$

	\section {Painlev\'e Analysis of the modified H\'enon-Heiles system}
	To start with we give a brief summary of the Painlev\'e analysis for a scalar $n^{th}$ order nonlinear ODE having the form \cite{Laksh}
	\begin{equation}\label{2.1}
		F\left(t, x,\frac{dx}{dt},....,\frac{d^nx}{dt^n} \right)=0,
	\end{equation}	
	where $F$ is analytic in $t$ and rational in other arguments. In the  Painlev\'e analysis, one looks for solution of equation (\ref{2.1}) expressed as a Laurent series
	\begin{equation}\label{2.2}
		x(t)=  (t-t_0)^{q_j}  \sum_{k}a_k (t-t_0)^k,~ ~ ~ ~ ~ Re(q_j)< 0,  ~ ~ ~ 0<|t-t_0|<R
	\end{equation}
	in the neighborhood of an arbitrary singular point $t_0 $. \\
	The Painlev\'e analysis for ODEs essentially consists of the following three steps:\\
	(a) determination of leading-order bahaviour of the Laurent series,\\
	(b) determination of resonances, that is the powers at which arbitrary constants of the solution of (\ref{2.2}) enter into the Laurent series expansion; and\\
	(c) verification that a sufficient number of arbitrary constants exist without the introduction of movable critical points.\\
	We would like to remind that the Painlev\'e analysis provides only necessary condition for integrability of the considered nonlinear ODE. The sufficient condition for integrability of it has to be established by other means, for example by constructing its sufficient number of integrals of motion.\\
	The remaining part of the section contains the details of the Painlev\'e analysis of the modified H\'enon-Heiles system governed by	
	
		\begin{equation}\label{1a}
			\frac{d^2 x}{dt^2}	 + \phi_1 \frac{dx}{dt} + A x + 2 \alpha x y  =0 ,
		\end{equation}
		
		\begin{equation}\label{1b}
			\frac{d^2 y}{dt^2}	 + \phi_2 \frac{dy}{dt} + B y + \alpha x^2 - \beta y^2 =0.
		\end{equation}

	\subsection{ Leading-Order behaviours }
	Let us assume that the leading order behaviour of $x(t)$ and $y(t)$ be
	\begin{equation}\label{5}
		x(t)\approx a_0 \tau^p, ~ y(t)\approx b_0 \tau^q, ~ \tau=(t-t_0)\rightarrow 0
	\end{equation}
	in a neighbourhood of the movable singularity $t_0$,
	which results to the leading-order equations	
	
		\begin{align}
			& a_0 p (p-1) \tau^{p-2} + 2\alpha a_0 b_0 \tau^{p+q} =0, \label{6a}\\
			& b_0 q(q-1)\tau^{q-2}+\alpha a_0^2\tau^{2p}-\beta b_0^2 \tau^{2q}=0.\label{6b}
		\end{align}
	
	From equations (\ref{6a}, \ref{6b}) we identify the following two distinct sets of possibilities.\\
	\textbf{Case 1}.
	\begin{equation}\label{7}
		p=-2, ~ q=-2,~ a_0=\pm \frac{3}{\alpha} \sqrt{\left(2+\frac{\beta}{\alpha}\right)}, ~b_0=-\frac{3}{\alpha}.
	\end{equation}
	\textbf{Case 2}.
	\begin{equation}\label{8}
		p=\frac{1}{2}\pm \frac{1}{2} \sqrt{ \left(1-\frac{48\alpha}{\beta}\right)} , ~q=-2,~ a_0=\mbox{arbitrary}, ~b_0=\frac{6}{\beta}.
	\end{equation}
	\subsection{Resonance}
	
	To identify the resonance values, we substitute
	\begin{equation}\label{9}
		x(t)\approx a_0 \tau^p+\omega_1 \tau^{p+r}, ~ y(t)\approx b_0\tau^q+\omega_2 \tau^{q+r} ~, \tau=(t-t_0)\rightarrow 0
	\end{equation}
	into equations (\ref{1a}) and (\ref{1b}), retaining only the leading-order terms.  This results into the following  system of linear algebraic equations,
	\begin{equation}\label{10}
		M_2(r)\Omega=0, ~\Omega=(\omega_1, \omega_2),
	\end{equation}
	where $M_2(r)$ is a $2\times2$ matrix depending on r.\\
	For Case 1, the form of $M_2(r)$ is
	\[M_{2}(r) =
	\left[ {\begin{array}{cc}
			(r-2)(r-3)+2\alpha b_0 & 2\alpha a_0 \\
			2\alpha a_0 & (r-2)(r-3)-2\beta b_0 \\
	\end{array} } \right].\]
	To have nontrivial solutions for $(\omega_1, \omega_2)$, we require $det(M_2(r))=0$,
	\begin{equation}\label{12}
		\Rightarrow	(r+1)(r-6)\left(r^2-5r+12+\frac{6\beta}{\alpha}\right)=0
	\end{equation}
	and so
	\begin{equation}\label{13}
		r=-1,~ 6, \frac{5}{2}\pm\frac{1}{2}\left[1-24\left(1+\frac{\beta}{\alpha}\right)\right]^{\frac{1}{2}}.
	\end{equation}
	\\
	In a similar manner for case 2 we find that the resonances occur at
	\begin{equation}\label{16}
		r=-1,~ 0,~ (1-2p),~ 6.
	\end{equation}
	Demanding that the resonance value $r$ be a non-negative  integer other than $-1$, equations (\ref{13}) and (\ref{16}) allow the following parametric cases and resonance values.
	
		\begin{align}	
			Case ~ 1(a):~&p = -2,~ ~ ~ q = -2; ~ ~ ~ \beta = -\alpha;~r=-1,2, 3,6.\label{16a}\\
			Case ~ 1(b):~&p = -2,~ ~ ~ q = -2; ~ ~ ~ \beta = -\frac{4}{3}\alpha;~r=-1,1, 4,6.\label{16b}\\
			Case ~ 1(c):~&p = -2,~ ~ ~ q = -2; ~ ~ ~ \beta = -2\alpha;~r=-1,0, 5,6.\label{16c}
		\end{align}
	
		\begin{align}
			Case ~ 2(a):~&p = -\frac{3}{2},~ ~ ~ q = -2; ~ ~ ~ \beta = -\frac{16}{5}\alpha;~r=-1,0,4,6.\label{17a}\\
			Case ~ 2(b):~&p = -1,~ ~ ~ q = -2; ~ ~ ~ \beta = -6\alpha; ~ ~r=-1,0,3,6.\label{17b}\\
			Case ~ 2(c):~&p = -\frac{1}{2},~ ~ ~ q = -2; ~ ~ ~ \beta = -16\alpha;~r=-1,0,2,6.\label{17c}		
		\end{align}
	
	\subsection{Evaluation of arbitrary constants}
	Obviously the movable singular point $t_0$ is arbitrary, associated with resonance value $-1$.  To check if the appropriate constants are arbitrary corresponding to other resonance values, we introduce the following series expansions,
	\begin{equation}\label{20a,b}
		x(t)= \sum_{k=0}^{6} a_k \tau^{p+k},~~~~ y(t)=  \sum_{k=0}^{6} b_k \tau^{q+k}, ~ ~ \tau = (t-t_0),
	\end{equation}
	in all the terms of (\ref{1a}) and (\ref{1b}). As a result, we obtain a system of $~2-$ coupled recurrence relations. For example, for case $1$, the recurrence relations read
	\begin{equation}\label{21a}
		(k-2)(k-3)a_k + (k-3)\phi_1a_{k-1} + Aa_{k-2} +2\alpha \sum_l a_{l}b_{k-l}=0,
	\end{equation}
	\begin{equation}\label{21b}
		(k-2)(k-3)b_k + (k-3)\phi_2b_{k-1} + Bb_{k-2} + \alpha \sum_l a_la_{k-l} -\beta \sum_lb_lb_{k-l}=0.
	\end{equation}

	Solving the above set of recurrence relations successively one can obtain the various $a_k$ and $b_k$ explicitly. For example, for  case 1(a),  when $k=0,1$ respectively we find $[Here, a_{-1}=b_{-1}=a_{-2}=b_{-2}=0]$
	\begin{equation}
		a_0=b_0=-\frac{3}{\alpha},~~
		a_1 = -\frac{3}{5\alpha}(2\phi_1-3\phi_2), ~~ b_1 = \frac{3}{5\alpha}(3\phi_1-2\phi_2).
	\end{equation}
	For $k=2$ we have
	
		\begin{align}
			&2\alpha(b_0a_2+a_0b_2)=(\phi_1-2\alpha b_1)a_1 -Aa_0, \\ &2\alpha(a_0a_2+b_0b_2)=-\alpha(a_1^2+b_1^2)+\phi_2b_1-Bb_0.
		\end{align}
	
	Solving the above equations we find

		\begin{align}
			&a_2 = arbitrary,~
			b_2 = -\frac{25A+26\phi_1^2-63\phi_1\phi_2+36\phi_2^2+50a_2\alpha}{50\alpha},\\
			&B = A-6\phi_1\phi_2+\frac{17}{5}\phi_2^2+\frac{13}{5}\phi_1^2.
		\end{align}
	
	For $k=3$ we have
	
		\begin{align}
			&2\alpha(b_0a_3+a_0b_3)=-2\alpha(a_1b_2+a_2b_1) -Aa_1,\\ &2\alpha(a_0a_3+b_0b_3)=-2\alpha(a_1a_2+b_1b_2)-Bb_1.
		\end{align}
	
	Solving the above equations we find
	\begin{equation}
		a_3=-b_3+\frac{\phi_1^3}{250\alpha},~b_3 = arbitrary,~\phi_2=\phi_1.
	\end{equation}
	Proceeding further we find $a_4, b_4, a_5$ and $b_5$ explicitly. However the coefficient either $a_6$ or $b_6$ is arbitrary only if
	\begin{equation}\label{25}
		B=A= \frac{6}{25} \phi_{{1}}^2.
	\end{equation}
	We thus infer that the modified H\'enon-Heiles system admits a single valued solution with four arbitrary constants for the parametric restrictions
	\begin{equation}\label{26}
		\beta=-\alpha,~	\phi_{{2}}=\phi_{{1}}=\phi,~A= B= \frac{6}{25} \phi^2 .
	\end{equation}
	
	\noindent Proceeding in a similar manner we find that for the following restrictions the modified H\'enon-Heiles system admits Laurent series solutions with four arbitrary constants:
	
	\begin{equation}\label{30}
		\noindent Case~2(b): ~ ~ A=B,~	\beta = -6 \alpha,~	\phi_{{1}}=\phi_{{2}}=\phi,~ 	B= \frac{6}{25} \phi^2,
	\end{equation}
	\begin{equation}\label{31}
		\noindent Case~2(c): ~ ~ A=B,~ \beta = -16 \alpha, ~ \phi_{{1}}=\phi_2=\phi,\, ~ B=\frac{6}{25} \phi^2,
	\end{equation}	
	We would like to mention that for the remaining cases $(1b),(1c)$ and $(2a)$ there exists a Laurent series solution with only three or two arbitrary constants and consequently special exact solutions may exist for these cases.\\
\subsection{Remark:}We also mention here the fact that on taking the condition  $\phi_1=\phi_2=\phi$ in \eqref{1a} and \eqref{1b} it leads directly to Painlev\'e integrability as follows: If one changes the variables $x,y$ to $x_1, x_2$ such that
\begin{equation}
    x=x_1e^{-\frac{\phi}{2} t} ,~~y=x_2e^{-\frac{\phi}{2} t},  
\end{equation}
one obtains from \eqref{1a} and \eqref{1b} a system of equations without a friction term where the exponential dependence on time only enters multiplicatively in the nonlinear terms,
\begin{equation}\label{&}
    \frac{d^2 x_1}{dt^2}=\left(\frac{\phi^2}{4}-A\right)x_1-2\alpha x_1 x_2 e^{-\frac{\phi}{2} t},~~ \frac{d^2 x_2}{dt^2}=\left(\frac{\phi^2}{4}-B\right)x_2-\left(\alpha x_1^2-\beta x_2^2\right)e^{-\frac{\phi}{2}t}. 
\end{equation}
Then, the Painlev\'e analysis directly applies to \eqref{&} above to show that all solutions are Laurent series, since a Taylor series expansion of $e^{-\frac{\phi}{2} t}$ would not affect the lower order calculations, where free constants enter, and the Painlev\'e property is established.

\section{The Prelle-Singer method}
	Given a scalar or coupled second order ODE there exist methods in the literature to find its integrals of motion, if they exist. In this article, we restrict our attention to the Prelle-Singer method. To start with we provide a brief account of the Prelle-Singer method for coupled second order ODEs \cite{Vkc1,Vkc2,Pre}.
	\subsection{The Prelle-Singer method for two-coupled second order ODEs }
	Consider a system of two-coupled second order ODEs having the form

\begin{align}
\noindent \ddot{x}&=\frac{P_1(t,x,y,\dot{x},\dot{y})}{Q_1(t,x,y,\dot{x},\dot{y})}=\Psi_1(x,y,\dot{x},\dot{y}),\label{32a}\\
\ddot{y}&=\frac{P_2(t,x,y,\dot{x},\dot{y})}{Q_2(t,x,y,\dot{x},\dot{y})}=\Psi_2(x,y,\dot{x},\dot{y})\label{32b}.
\end{align}

\noindent where $\dot{x}=\frac{dx}{dt},~\dot{y}=\frac{dy}{dt},~\ddot{x}=\frac{d^2x}{dt^2},~\ddot{y}=\frac{d^2y}{dt^2}, ~ P_i,~Q_i,~i=1,2 $ are polynomials in $(t,~ x,~ y, ~\dot{x}, \dot{y})$ with real or complex coefficients.\\
Let us assume that the system (\ref{32a}) and (\ref{32b})  admits a first integral of the form
\begin{equation}\label{33}
I(t,~x,~y,~\dot{x},~\dot{y})=C ,
\end{equation}  where C is constant and so
\begin{equation} \label{34a}
dI=I_t dt +I_x dx +I_y dy +I_{\dot{x}} d{\dot{x}}\, +I_{\dot{y}}\, d\dot{y}=0,
\end{equation} 	
where each subscript denotes partial derivative with respect to that variable. Equations (\ref{32a}) and (\ref{32b}) can be rewritten as

\begin{align}
\frac{P_1}{Q_1} dt\, - d\dot{x}\,=0,\label{34b}\\ \frac{P_2}{Q_2} dt\, - d\dot{y}\,=0.\label{34c}
\end{align}	

\noindent Adding a null term $S_1\, (t, x, y, \dot{x}, \dot{y} )\dot{x} dt - S_1\, (t, x, y, \dot{x}, \dot{y} ) dx$ and\\ $ ~S_2\, (t, x, y, \dot{x}, \dot{y} )\dot{y} dt - S_2\, (t, x, y, \dot{x}, \dot{y} ) dy$ to (\ref{34b}) and (\ref{34c}), respectively, we obtain  the following equations	

\begin{align}
(\Psi_1 dt\,+ S_1 \dot{x})dt-S_1 dx - d\dot{x}\,=0, \label{35a}\\
(\Psi_2 dt\,+S_2 \dot{y})dt-S_2 dy- d\dot{y}\,=0.\label{35b}
\end{align}	

\noindent Let $R_1(t, x, y, \dot{x}, \dot{y} )$ and $R_2(t, x, y, \dot{x}, \dot{y} )$ be the appropriate integrating factors for (\ref{35a}) and (\ref{35b}), respectively.
Hence we have
\begin{equation}\label{37}
dI=R_1\, (\Psi_1+S_1 \dot{x})dt+R_2\, (\Psi_2+S_2 \dot{y})dt-R_1 S_1 dx\,- R_2 S_2dy\,-R_1 d\dot{x}-R_2 d\dot{y}=0.
\end{equation}
On the solutions, we require that (\ref{34a}) and (\ref{37})  be proportional. Comparing equations (\ref{34a}) with (\ref{37}) we have,

\begin{align}
I_t&= R_1(\Psi_1 +S_1 \dot{x})+R_2 (\Psi_2+ S_2 \dot{y}),\label{38a}\\
I_x&= -R_1 S_1, ~I_y= -R_2 S_2, \label{38b}\\
I_{\dot{x}}&= -R_1 , ~I_{\dot{y}}=-R_2. \label{38c}
\end{align}

\noindent The compatibility conditions, 
 \begin{align*}
I_{tx} =I_{xt},\, I_{ty} =I_{yt},\,~I_{t\dot{x}}=I_{\dot{x}t} \,,~I_{t\dot{y}}=I_{\dot{y}t}\, ,~I_{xy} =I_{yx},
\\~I_{x\dot{x}}=I_{\dot{x}x},~ I_{x\dot{y}}=I_{\dot{y}x},\,~I_{y\dot{x}}=I_{\dot{x}y},\,~ I_{y\dot{y}}=I_{\dot{y}y}  ,   I_{\dot{x}\dot{y}}=I_{\dot{y}\dot{x}},\,
 \end{align*}
on the equations (\ref{38a}),~(\ref{38b}) and (\ref{38c}) leads to the following conditions,
	
\begin{align}
D[S_1]&=-\Psi_{1x}-\frac{R_2}{R_1}\,\Psi_{2x}+\frac{R_2}{R_1}\, S_1 \Psi_{2\dot{x}}+ S_1 \Psi_{1\dot{x}}+ S_1^2,\label{39}\\
D[S_2]&=-\Psi_{2y}-\frac{R_1}{R_2}\,\Psi_{1y}+\frac{R_1}{R_2}\, S_2 \Psi_{1\dot{y}}+ S_2 \Psi_{2\dot{y}}+ S_2^2,\label{40}\\
D[R_1]&=\frac{\partial R_1}{\partial{t}}+\dot{x} \frac{\partial R_1}{\partial{x}}+ \dot{y} \frac{\partial R_1}{\partial{y}}+\ddot{x} \frac{\partial R_1}{\partial{\dot{x}}}+ \ddot{y} \frac{\partial R_1}{\partial{\dot{y}}}\nonumber\\
&=-(R_1 \Psi_{1\dot{x}}+R_2 \Psi_{2\dot{x}}+R_1 S_1),\label{41}\\	
D[R_2]&=\frac{\partial R_2}{\partial{t}}+\dot{x} \frac{\partial R_2}{\partial{x}}+ \dot{y} \frac{\partial R_2}{\partial{y}}+\ddot{x} \frac{\partial R_2}{\partial{\dot{x}}}+ \ddot{y} \frac{\partial R_2}{\partial{\dot{y}}}\nonumber\\
&=-(R_2 \Psi_{2\dot{y}}+R_1 \Psi_{1\dot{y}}+R_2 S_2),\label{42}\\
S_1 R_{1y}&= -R_1 S_{1y} + S_2 R_{2x} +R_2 S_{2x},\label{43a}\\
R_{1x}&= S_1 R_{1\dot{x}}+R_1 S_{1\dot{x}}, R_{1y}= S_2 R_{2\dot{x}}+ R_2 S_{2\dot{x}} , \label{43}\\	
R_{2y}& = S_2 R_{2\dot{y}}+ R_2 S_{2\dot{y}},~	R_{2x}= S_1 R_{1\dot{y}}+ R_1 S_{1\dot{y}},\label{44}\\
R_{1\dot{y}}&=R_{2\dot{x}} ,\label{45}
\end{align}

\noindent where $D$ is the total differential operator given by
\begin{align*}
 D=\frac{\partial}{\partial{t}}+\dot{x} \frac{\partial}{\partial{x}}+ \dot{y} \frac{\partial}{\partial{y}}+\ddot{x} \frac{\partial}{\partial{\dot{x}}}+\ddot{y} \frac{\partial}{\partial{\dot{y}}}.
 \end{align*}

\noindent Integrating equations (\ref{38a}), (\ref{38b}) and (\ref{38c}) we obtain the corresponding integral of motion,
\begin{equation} \label{46}
I= \kappa_1 +\kappa_2 +\kappa_3 +\kappa_4- \int [R_2+ \frac{\partial}{\partial \dot{y}}(\kappa_1 +\kappa_2 +\kappa_3 +\kappa_4)] d\dot{y},
\end{equation}
where
\begin{equation*}
\kappa_1= \int \bigg(  R_1(\Psi_1 +S_1 \dot{x})+R_2 (\Psi_2+ S_2 \dot{y})\bigg) dt,
~~\kappa_2= -\int \bigg(R_1 S_1 +\frac{\partial}{\partial{x}}(\kappa_1) \bigg) dx,~~\\
\end{equation*}
\begin{equation*}
\kappa_3= -\int \bigg(R_2 S_2 +\frac{\partial}{\partial{y}}(\kappa_1+\kappa_2) \bigg) dy,~~
\kappa_4= -\int \bigg(R_1 +\frac{\partial}{\partial{\dot{x}}}(\kappa_1+\kappa_2 +\kappa_3) \bigg) d\dot{x}.
\end{equation*}

\noindent  The determining equations (\ref{39})-(\ref{45}) are highly coupled and cannot be solved directly. Hence we use the following method to solve it.\\
To start with, we can identify the following two identities using (\ref{39})-(\ref{42}),
\begin{equation} \label{47}
D[R_1 S_1]=D(R_1) S_1+R_1 D(S_1)=- (R_1 \Psi_{1 x}+ R_2 \Psi_{2x}),
\end{equation}
\begin{equation}\label{48}
D[R_2 S_2]=D(R_2) S_2+R_2 D(S_2)=- (R_1 \Psi_{1 y}+ R_2 \Psi_{2y}).
\end{equation}

To eliminate $S_1$ and $S_2$ we again take total derivative of equations (\ref{41}) and (\ref{42}) with respect to $'t'$ and then using equations (\ref{47}) and (\ref{48})  we get
\begin{equation}
\begin{aligned}\label{49}
& R_{1 t t}+2 \dot{x} R_{1 t x}+2 \dot{y} R_{1 t y}+2 \Psi_1 R_{1 t \dot{x}}+2 \Psi_2 R_{1 t \dot{y}}+\dot{x}^2 R_{1 x x}+2 \dot{x} \dot{y} R_{1 x y} \\
& +\dot{y}^2 R_{1 y y}+\Psi_{1 t} R_{1 \dot{x}}+\Psi_{2 t} R_{1 \dot{y}}+\dot{x} \Psi_{1 x} R_{1 \dot{x}}+\dot{y} \Psi_{1 y} R_{1 \dot{x}}+2 \dot{x} \Psi_1 R_{1 x \dot{x}}\\
&+2 \dot{y} \Psi_1 R_{1 y \dot{x}}	+2 \dot{x} \Psi_2 R_{1 x \dot{y}}+2 \dot{y} \Psi_2 R_{1 y \dot{y}}
+\dot{x} \Psi_{2 x} R_{1 \dot{y}}+\dot{y} \Psi_{2 y} R_{1 \dot{y}}+\Psi_1 R_{1 x}\\
&+\Psi_2 R_{1 y}+\Psi_1 \Psi_{1 \dot{x}} R_{1 \dot{x}}+\Psi_1^2 R_{1 \dot{x} \dot{x}}+\Psi_2 \Psi_{1 \dot{y}} R_{1 \dot{x}}+\Psi_1 \Psi_{2 \dot{x}} R_{1 \dot{y}}+\Psi_2 \Psi_{2 \dot{y}} R_{1 \dot{y}}\\
&	+\Psi_{1 \dot{x}}\left(R_{1 t}+\dot{x} R_{1 x}+\dot{y} R_{1 y}+\Psi_1 R_{1 \dot{x}}+\Psi_2 R_{1 \dot{y}}\right)+\Psi_2^2 R_{1 \dot{y} \dot{y}}+2 \Psi_1 \Psi_2 R_{1 \dot{x} \dot{y}}\\
&  +\Psi_{2 \dot{x}}\left(R_{2 t}+\dot{x} R_{2 x}+\dot{y} R_{2 y}+\Psi_1 R_{2 \dot{x}}+\Psi_2 R_{2 \dot{y}}\right) -R_1 \Psi_{1 x}-R_2 \Psi_{2 x}\\
&+R_1\left(\Psi_{1 t \dot{x}}+\dot{x} \Psi_{1 x \dot{x}}+\dot{y} \Psi_{1 y \dot{x}}+\Psi_1 \Psi_{1 \dot{x} \dot{x}}+\Psi_2 \Psi_{1 \dot{x} \dot{y}}\right)\\
& +R_2(\Psi_{2 t \dot{x}}+\dot{x} \Psi_{2 x \dot{x}}+\dot{y} \Psi_{2 y \dot{x}}+\Psi_1 \Psi_{2 \dot{x} \dot{x}}+\Psi_2 \Psi_{2 \dot{x} \dot{y}})=0,
\end{aligned}
\end{equation}

\begin{equation}
\begin{aligned}\label{50}
& R_{2 t t}+2 \dot{x} R_{2 t x}+2 \dot{y} R_{2 t y}+2 \Psi_1 R_{2 t \dot{x}}+2 \Psi_2 R_{2 t \dot{y}}+\dot{x}^2 R_{2 x x}+2 \dot{x} \dot{y} R_{2 x y}+\dot{y}^2 R_{2 y y} \\
&  +\Psi_{1 t} R_{2 \dot{x}}	+\Psi_{2 t} R_{2 \dot{y}}+\dot{x} \Psi_{1 x} R_{2 \dot{x}}+\dot{y} \Psi_{1 y} R_{2 \dot{x}}+2 \dot{x} \Psi_1 R_{2 x \dot{x}}+2 \dot{y} \Psi_1 R_{2 y \dot{x}}	\\
& +2 \dot{x} \Psi_2 R_{2 x \dot{y}}+2 \dot{y} \Psi_2 R_{2 y \dot{y}}
+\dot{x} \Psi_{2 x} R_{2 \dot{y}}+\dot{y} \Psi_{2 y} R_{2 \dot{y}}+\Psi_1 R_{2 x}+\Psi_2 R_{2 y}\\
& +\Psi_1 \Psi_{1 \dot{x}} R_{2 \dot{x}}	+\Psi_1^2 R_{2 \dot{x} \dot{x}}+\Psi_2 \Psi_{1 \dot{y}} R_{2 \dot{x}}
+\Psi_1 \Psi_{2 \dot{x}} R_{2 \dot{y}}+\Psi_2 \Psi_{2 \dot{y}} R_{2 \dot{y}}\\
& +\Psi_{1 \dot{y}}\left(R_{1 t}+\dot{x} R_{1 x}+\dot{y} R_{1 y}+\Psi_1 R_{1 \dot{x}}+\Psi_2 R_{1 \dot{y}}\right)
+\Psi_2^2 R_{2 \dot{y} \dot{y}}\\
& +2 \Psi_1 \Psi_2 R_{2 \dot{x} \dot{y}}+\Psi_{2 \dot{y}}\left(R_{2 t}+\dot{x} R_{2 x}+\dot{y} R_{2 y}+\Psi_1 R_{2 \dot{x}}+\Psi_2 R_{2 \dot{y}}\right)\\
& -R_1 \Psi_{1 y}-R_2 \Psi_{2 y}+R_1\left(\Psi_{1 t \dot{y}}+\dot{x} \Psi_{1 x \dot{y}}+\dot{y} \Psi_{1 y \dot{y}}+\Psi_1 \Psi_{1 \dot{x} \dot{y}}+\Psi_2 \Psi_{1 \dot{y} \dot{y}}\right)\\
& +R_2\left(\Psi_{2 t \dot{y}}+\dot{x} \Psi_{2 x \dot{y}}+\dot{y} \Psi_{2 y \dot{y}}+\Psi_1 \Psi_{2 \dot{x} \dot{y}}+\Psi_2 \Psi_{2 \dot{y} \dot{y}}\right)=0.
\end{aligned}
\end{equation} 

\noindent Also differentiating equation (\ref{41}) with respect to $\dot{x}$ and equation (\ref{42}) with respect to $\dot{y}$, we have
\begin{equation}
\begin{aligned}	\label{52}
&R_{1t\dot{x}}+\,\dot{x}R_{1x\dot{x}} + \dot{y} R_{1y\dot{x}}+ \Psi_1 R_{1\dot{x} \dot{x}} +\Psi_2 R_{2\dot{x}\dot{x}}+ 2 R_{1x}+2\Psi_{2\dot{x}} R_{2\dot{x}} \\
&+2\Psi_{1 \dot{x}}R_{1\dot{x}}+R_2 \Psi_{2\dot{x}\dot{x}}+ R_1 \Psi_{1 \dot{x} \dot{x}}  =0,
\end{aligned}
\end{equation}
	
\begin{equation}
\begin{aligned}	\label{53}
&R_{2t\dot{y}}+\,\dot{x}R_{2x\dot{y}} + \dot{y} R_{2y\dot{y}}+ \Psi_1 R_{1\dot{y} \dot{y}} +\Psi_2 R_{2\dot{y}\dot{y}}+ 2 R_{2y}+2\Psi_{2\dot{y}} R_{2\dot{y}} \\
&+2\Psi_{1 \dot{y}}R_{1\dot{y}}+ R_2 \Psi_{2\dot{y}\dot{y}}+R_1 \Psi_{1 \dot{y} \dot{y}}  =0.
\end{aligned}
\end{equation}
	
\noindent Differentiating (\ref{41}) with respect to $\dot{y}$ or (\ref{42}) with respect to $\dot{x}$ and using  (\ref{43}) and (\ref{44}), we get a single equation
\begin{equation}
\begin{aligned}	\label{54}
&R_{1t\dot{y}}+\,\dot{x}R_{1x\dot{y}} + \dot{y} R_{1y\dot{y}}+ \Psi_1 R_{1\dot{x} \dot{y}} +\Psi_2 R_{1\dot{y}\dot{y}}+ R_{1y}+R_{2x}+\Psi_{2\dot{x}} R_{2\dot{y}} \\
&+\Psi_{1 \dot{x}}R_{1\dot{y}}+ \Psi_{2\dot{x}\dot{y}} R_2 + \Psi_{1 \dot{x} \dot{y}} R_1+ \Psi_{1\dot{y}}R_{1\dot{x}}+\Psi_{2\dot{y}} R_{1\dot{y}} =0.
\end{aligned}
\end{equation}

Using equations (\ref{43}) ,(\ref{44}), (\ref{52}), (\ref{53}) and (\ref{54}), equation (\ref{49}) simplifies as

\begin{equation}
\begin{aligned} \label{55}
&R_{1 t t}+2 \dot{x} R_{1 t x}+2 \dot{y} R_{1 t y}+\Psi_1 R_{1 t \dot{x}}+\Psi_2 R_{1 t \dot{y}}+\dot{x}^2 R_{1 x x}+2 \dot{x} \dot{y} R_{1 x y}-R_1 \Psi_{1 x} \\
&-R_2 \Psi_{2 x}+\dot{y}^2 R_{1 y y}+\Psi_{1 t} R_{1 \dot{x}}+\Psi_{2 t} R_{1 \dot{y}}+\dot{x} \Psi_{1 x} R_{1 \dot{x}}+\dot{y} \Psi_{1 y} R_{1 \dot{x}}+\dot{x} \Psi_1 R_{1 x \dot{x}} \\
&+\dot{y} \Psi_1 R_{1 y \dot{x}}+\dot{x} \Psi_2 R_{1 x \dot{y}}+\dot{y} \Psi_2 R_{1 y \dot{y}}+\dot{x} \Psi_{2 x} R_{1 \dot{y}}+\dot{y} \Psi_{2 y} R_{1 \dot{y}}-\Psi_1 R_{1 x} \\
&-\Psi_2 R_{2 x}+\Psi_{1 \dot{x}}\left(R_{1 t}+\dot{x} R_{1 x}+\dot{y} R_{1 y}\right)+R_1\left(\Psi_{1 t \dot{x}}+\dot{x} \Psi_{1 x \dot{x}}+\dot{y} \Psi_{1 y \dot{x}}\right) \\
&+\Psi_{2 \dot{x}}\left(R_{2 t}+\dot{x} R_{2 x}+\dot{y} R_{2 y}\right)+R_2\left(\Psi_{2 t x}+\dot{x} \Psi_{2 x \dot{x}}+\dot{y} \Psi_{2 y \dot{y}}\right)=0,
\end{aligned}
\end{equation}
Now, using equations (\ref{43}) ,(\ref{44}), (\ref{52}), (\ref{53}) and (\ref{54}) , equation (\ref{50}) simplifies to the form
\begin{equation}
\begin{aligned}\label{56}
&R_{2 t t}+2 \dot{x} R_{2 t x}+2 \dot{y} R_{2 t y}+\Psi_1 R_{2 t \dot{x}}+\Psi_2 R_{2 t \dot{y}}+\dot{x}^2 R_{2 x x}+2 \dot{x} \dot{y} R_{2 x y}-R_1 \Psi_{1 y} \\
&-R_2 \Psi_{2 y}+\Psi_{1 t} R_{2 \dot{x}}+\Psi_{2 t} R_{2 \dot{y}}+\dot{y}^2 R_{2 y y}+\dot{x} \Psi_{1 x} R_{2 \dot{x}}+\dot{y} \Psi_{1 y} R_{2 \dot{x}}+\dot{x} \Psi_1 R_{2 x \dot{x}} \\
&+\dot{y} \Psi_1 R_{2 y \dot{x}}+\dot{x} \Psi_2 R_{2 x \dot{y}}+\dot{y} \Psi_2 R_{2 y \dot{y}}+\dot{x} \Psi_{2 x} R_{2 \dot{y}}+\dot{y} \Psi_{2 y} R_{2 \dot{y}}-\Psi_1 R_{1 y} \\
&-\Psi_2 R_{2 y}+\Psi_{1 \dot{y}}\left(R_{1 t}+\dot{x} R_{1 x}+\dot{y} R_{1 y}\right)+R_1\left(\Psi_{1 t \dot{y}}+\dot{x} \Psi_{1 x \dot{y}}+\dot{y} \Psi_{1 y \dot{y}}\right) \\
&+\Psi_{2 \dot{y}}\left(R_{2 t}+\dot{x} R_{2 x}+\dot{y} R_{2 y}\right)+R_2\left(\Psi_{2 t \dot{y}}+\dot{x} \Psi_{2 x y}+\dot{y} \Psi_{2 y \dot{y}}\right)=0.
\end{aligned}
\end{equation}
	
\noindent  Now solving the equations (\ref{55}) and (\ref{56}) one can find $R_1$ and $R_2$. The null forms $S_1$ and $S_2$ can be found by substituting the obtained values of $R_1$ and $R_2$ in equations (\ref{41}) and (\ref{42}).
Hence integrals of motion for the system (\ref{32a}) and (\ref{32b}) can be found by using the resultant forms of $R_1,~ R_2,~ S_1$ and $S_2$ in equation (\ref{46}).

\subsection{The Prelle-Singer method for modified H\'enon-Heiles system}
	
\noindent We rewrite the modified H\'enon-Heiles system  given by equations (\ref{1a}) and (\ref{1b}) for $\phi_1=\phi_2=\phi$ as

\begin{align}
\ddot{x}	&=  \Psi_1 (x,y,\dot{x})=-\phi \dot{x} - A x - 2 \alpha x y  , \label{4a}\\
\ddot{y}	&  =\Psi_2 (x,y,\dot{y})= - \phi \dot{y} - B y - \alpha x^2 + \beta y^2. \label{4b}
\end{align}

Let us assume that the above equations admit an integral of motion $I(t,~x,~y,~\dot{x},~\dot{y}).$ Following the procedure outlined in the previous sub-section we get
	
\begin{equation}
\begin{aligned}\label{61aa}
&	R_{1 t t}+2 \dot{x} R_{1 t x}+2 \dot{y} R_{1 t y}+\Psi_1
R_{1 t \dot{x}}+\Psi_2 R_{1 t \dot{y}}+\dot{x}^2 R_{1 x x}+2 \dot{x} \dot{y} R_{1 x y}-\Psi_2 R_{2 x}\\
&	+(A+2\alpha y) R_1+2\alpha x R_2 +\dot{y}^2 R_{1 y y}-\phi \dot{x} R_{1 \dot{x}}-2\alpha x\dot{y} R_{1 \dot{x}}+\dot{x} \Psi_1 R_{1 x \dot{x}}\\
&	+\dot{y} \Psi_1 R_{1 y \dot{x}}+\dot{x} \Psi_2 R_{1 x \dot{y}}+\dot{y} \Psi_2 R_{1 y \dot{y}}
-2 \alpha x \dot{x} R_{1 \dot{y}}-\Psi_1 R_{1 x} \\
&	+ (2\beta y-B)\dot{y} R_{1 \dot{y}}-\phi \left(R_{1 t}+\dot{x} R_{1 x}+\dot{y} R_{1 y}\right)=0,
\end{aligned}
\end{equation}
\begin{equation}
\begin{aligned}\label{62}
&	R_{2 t t}+2 \dot{x} R_{2 t x}+2 \dot{y} R_{2 t y}+\Psi_1 R_{2 t \dot{x}}+\Psi_2 R_{2 t \dot{y}}+\dot{x}^2 R_{2 x x}+2 \dot{x} \dot{y} R_{2 x y} +2\alpha x R_1 \\
&	+(B+2\beta y)R_2 -(A+2\alpha y)\dot{x} R_{2 \dot{x}}-2\alpha x \dot{y} R_{2 \dot{x}}+\dot{x} \Psi_1 R_{2 x \dot{x}} + \Psi_1 \dot{y} R_{2 y \dot{x}}\\
&	+\dot{y}^2 R_{2 y y}+\dot{x} \Psi_2 R_{2 x \dot{y}}+\dot{y} \Psi_2 R_{2 y \dot{y}}-(B-2\beta y)\dot{y} R_{2 \dot{y}}-\Psi_1 R_{1 y} \\
&	-2\alpha x \dot{x}  R_{2 \dot{y}}-\Psi_2 R_{2 y}-\phi \left(R_{2 t}+\dot{x} R_{2 x}+\dot{y} R_{2 y}\right)=0.
\end{aligned}
\end{equation}
\noindent A comprehensive examination reveals that obtaining a general solution for the aforementioned equations with respect to $R_1$ and $R_2$ is generally unattainable. Nonetheless, a specific solution is sufficient for determining the integrals of motion. Therefore, we assume that \begin{equation*}
\begin{aligned}
& R_1 = [(\epsilon_{11} x + \epsilon_{12}y+\epsilon_{13})\dot{x}+(\lambda_{11} x + \lambda_{12}y+\lambda_{13})\dot{y}+(\mu_{11} x + \mu_{12}y+\mu_{13})]f(t),\\
& R_2 = [(\epsilon_{21} x + \epsilon_{22}y+\epsilon_{23})\dot{x}+(\lambda_{21} x + \lambda_{22}y+\lambda_{23})\dot{y}+(\mu_{21} x + \mu_{22}y+\mu_{23})]f(t).
\end{aligned}
\end{equation*}
where $\epsilon_{ij},~\lambda_{ij}, ~\mu_{ij}, ~ i = 1,2,  j=1,2,3 $ are unknown constants and $f(t)$ is an unknown function of $t$ which has to be determined.\\ Substituting the values of $R_1$ and $R_2$ into (\ref{61aa}) and (\ref{62}), solving the resulting equations we can obtain the following two sets of values for case 1(a).

\begin{equation*}
\begin{aligned}
&(i)~\lambda_{1j}=\epsilon_{2j}=\lambda_{2k} =\epsilon_{1k}=\mu_{12}=\mu_{13}=0,~k=1,2,~ j=1,2,3.\\
&~\epsilon_{13}=-1,~\lambda_{23}=-1, \mu_{11}=\mu_{21}=-\frac{2}{5}\phi,\mu_{22}=\mu_{23}=0,~f(t)=e^{\frac{6}{5}\phi t}.\\
&(ii) ~\epsilon_{1j}=\epsilon_{2k}=\lambda_{1k}=\lambda_{2j}=\mu_{11}=\mu_{13}=0, ~k=1,2, ~j=1,2,3.\\
&\mu_{22}=\mu_{23}=0, ~\lambda_{13}=\epsilon_{23}=-1,~\mu_{12}=\mu_{21}=-\frac{2}{5}\phi, f(t)=e^{\frac{6}{5}\phi t}.
\end{aligned}
\end{equation*}
Below we give the explicit forms of $R_1$, $R_2$, $S_1$, $S_2$ and their integrals of motion for case (1a).\\
\textbf{Case 1(a).}$ ~A=B,~\beta= -\alpha,  ~B=\frac{6}{25} \phi^2 .$\\	
Here the explicit forms  of $R_1$ and $R_2$ read
\begin{equation*}
\begin{aligned}
(i) ~R_1 &=-\left(\dot{x}+\frac{2}{5}\,\phi x\right) e^{\frac{6}{5}\phi t } ,
~~~ &R_2=-\left(\dot{y}+\frac{2}{5}\,\phi y\right) e^{\frac{6}{5}\phi t },\\(ii) ~ R_1 &=-\left(\dot{y}+\frac{2}{5}\,\phi y\right) e^{\frac{6}{5}\phi t }  ,
~~~&R_2=-\left(\dot{x}+\frac{2}{5}\,\phi x\right) e^{\frac{6}{5}\phi t }, \\
\end{aligned}
\end{equation*}
	
\noindent Making use of the above $R_1$ and $R_2$ in equations (\ref{41}) and (\ref{42}) we obtain the following values of $S_1$ and $S_2$ respectively.
\begin{equation*}
\begin{aligned}
(i)~ S_1&=\frac{2}{5}\,\phi +\,{\frac {2\alpha\,xy}{\dot{x} +\frac{2}{5}\phi x}},~~
&S_2=\frac{2}{5}\,\phi +{\frac {\alpha\, \left( {x}^{2}+\,{y}^{2} \right) }{\dot{y} +\frac{2}{5}\,\phi y}},\\
(ii)  ~S_1&=\frac{2}{5}\phi +{\frac {\alpha\,({y}^{2}+{x}^{2})}{\dot{y}+\frac{2}{5}\phi  \,y}},
~~&S_2=\frac{2}{5}\,\phi +2\alpha\,\,{\frac {xy}{\dot{x}+\frac{2}{5}\phi x}},	
\end{aligned}
\end{equation*}
	
\noindent Thus we  obtain the following integrals of motion for (\ref{4a}) and (\ref{4b}),

\begin{equation*}
\begin{aligned}	
\noindent (i) ~I_1&=\Bigg(\frac{1}{2}\Big(\dot{x}^2+\dot{y}^2\Big)+\frac{2}{5} \phi\Big(x\dot{x}+y\dot{y}\Big)+\frac{2}{25}\phi^2\Big(x^2+y^2\Big)+\alpha y \Big(x^2+\frac{1}{3}y^2\Big)\Bigg) e^{\frac{6}{5}\,\phi\,t} ,\label{65a}\\
(ii)~{I_2}&=\left(\dot{x}\,\dot{y}+\frac{2}{5}\phi\Big(\dot{x}y+x\dot{y}\Big)+\frac{4}{25}\phi^2 xy+\alpha x\Big(y^2+\frac{1}{3}x^2\Big)\right)\,e^{\frac{6}{5}\,\phi\,t} .
\end{aligned}
\end{equation*}
	
Proceeding further we obtain nontrivial  forms for $R_1$ and $R_2$ associated with  case 2(b).\\
The explicit forms of $R_1,~R_2,~S_1 $,~$S_2$ and the integrals of motion obtained for the parametric restrictions in case 2(b) are given below.\\
\textbf{Case 2(b).}	$~ A=B,~\beta= -6\alpha,~ B=\frac{6}{25} \phi^2. $\\
The explicit forms of $R_1,R_2,S_1$ and $S_2$ and the related integrals read as\\	
\begin{equation*}
\begin{aligned}
(i) ~R_1&=-(\dot{x}+\frac{2}{5}\,\phi x) e^{\frac{6}{5}\phi t },
~~R_2 =-(\dot{y}+\frac{2}{5}\,\phi y) e^{\frac{6}{5}\phi t },\\
S_1&=\frac{2}{5}\,\phi +\,{\frac {2\alpha\,xy}{\dot{x} +\frac{2}{5}\phi x}},
~~~S_2 =\frac{2}{5}\,\phi +{\frac {\alpha\, \left( {x}^{2}+6\,{y}^{2} \right) }{\dot{y} +\frac{2}{5}\,\phi y}}.\\
I_1&=\Bigg(\frac{1}{2}\Big(\dot{x}^2+\dot{y}^2\Big)+\frac{2}{5}\phi\Big(x\dot{x}+y\dot{y}\Big)+\frac{2}{25}\phi^2\Big(x^2+y^2\Big)+\alpha y\Big(x^2+2y^2\Big)\Bigg) e^{\frac{6}{5} \phi  t},\\		(ii)~R_1&=4(2\dot{x}y-x\dot{y}+\frac{2}{5}\,\phi xy) e^{\frac{8}{5}\phi t },
~~R_2 =-4x(\dot{x}+\frac{2}{5}\,\phi x) e^{\frac{8}{5}\phi t }, \\
~S_1&=\dfrac{\dot{x} \dot{y}+\frac{4}{5}\,\phi x\, \dot{y}+{\frac {4}{25}}\,{\phi }^{2}xy+\alpha\,{x}^{3}+2\,
\alpha\,x{y}^{2}} {x \dot{y}- 2\dot{x} y-\frac{2}{5} \phi xy},\\
S_2&=\dfrac{-(\dot{x}^2+\frac{2}{5} \phi x \dot{x}- 2 \alpha x^2 y)}{x \dot{y}+\frac{2}{5} \phi  x^2},\\		
{I_2}&= \Bigg(4\dot{x} \Big(x\dot{y}-y\dot{x}\Big)+\frac{8}{5}\phi x\Big(x \dot{y}-\dot{x}y\Big)+\alpha x^2\Big(x^2+4y^2\Big)\Bigg) e^{\frac{8}{5}\phi  t}.\label{66b}		
\end{aligned}
\end{equation*}	
It is easy to check that the obtained integrals are independent in both the cases.\\
In order to construct the integral for case 2(c) we consider $R_1$ and $R_2$ as quadratic polynomials in $\dot{x}$ and $\dot{y}$ as
$$ R_1 = [(\epsilon_1 x + \epsilon_2y+\epsilon_3)\dot{x}^2+(\epsilon_4 x + \epsilon_5y+\epsilon_6)\dot{y}^2+(\epsilon_7 x + \epsilon_8y+\epsilon_9)\dot{x}\dot{y}+\cdots]f(t)$$
$$ R_2 = [(\theta_1 x + \theta_2y+\theta_3)\dot{x}^2+(\theta_4 x + \theta_5y+\theta_6)\dot{y}^2+(\theta_7 x + \theta_8y+\theta_9)\dot{x}\dot{y}+\cdots]f(t),$$
where $\epsilon_i, \theta_i,~ i= 1,2,...$ are constants and $f(t)$ is an unknown function. A detailed calculation shows that there exists  nontrivial solutions for $R_1$, $R_2$, $S_1$ and $S_2$  in this case also. Their explicit forms  for the case $2(c)$ are given below.\\
\noindent \textbf{Case 2(c).}	$ A=B,~\beta= -16\alpha,  ~B=\frac{6}{25} \phi^2 .$	
\begin{equation*}
\begin{aligned}
(i) ~R_1 &=- (\dot{x}+\frac{2}{5}\,\phi x )  e^{\frac{6}{5}\,\phi t},
~ R_2=-(\dot{y}+\frac{2}{5}\,\phi y) e^{\frac{6}{5}\phi t },\\
S_1 &=\frac{2}{5}\,\phi +\,{\frac {2\alpha\,xy}{\dot{x}+\dfrac{2}{5}\,\phi x}} ,~~
S_2 =\frac{2}{5}\,\phi +{\frac {\alpha\, \left( {x}^{2}+16\,{y}^{2} \right) }{
\dot{y}+\frac{2}{5}\,\phi y}},	\\
~I_1&=\Bigg(\frac{1}{2}\Big(\dot{x}^2+\dot{y}^2\Big)+\frac{2}{5}\phi\Big(x\dot{x}+y\dot{y}\Big)+\frac{2}{25} \phi^2 \Big(x^2+y^2\Big)+\alpha y\Big(x^2+\frac{16}{3}y^2\Big)\Bigg) e^{\frac{6}{5} \phi t},\\	
(ii) ~R_1&=-\Bigg(36\,{\dot{x}}^{3}+{\frac {216}{5}}\,{\dot{x}}^{2}\phi x+{\frac {432}{25}}\,\dot{x}{
\phi }^{2}{x}^{2}+{\frac {288}{125}}\,{\phi }^{3}{x}^{3}+72
\,\alpha\,{x}^{2}y\dot{x}
+24\,\alpha\,{x}^{3}\phi y-12\,\alpha\,{x}^{3}\dot{y}\Bigg) e^{\frac{12}{5}\phi t} ,\\& ~R_2=12\alpha\,x^3\,(\dot{x}+\frac{2}{5}\,\phi x) e^{\frac{12}{5}\phi t }, ~S_1= \frac{U(x,y,\dot{x},\dot{y})}{V(x,y,\dot{x},\dot{y})} , ~~~~~S_2= {\frac {3\,\dot{x}^2+2\,\alpha\phi\,x\dot{x}-4\alpha x^2 y+\frac{8}{25}\,\phi^2 x^2}{x(\dot{x}+\dfrac{2}{5}x)}}.
\end{aligned}
\end{equation*}
where 	
\begin{equation*}
\begin{aligned}
U(x,y,\dot{x},\dot{y})&=\frac{72}{5} \dot{x}^2\Big(\dot{x}+\frac{6}{5}\phi x\Big)+72 \alpha xy \dot{x}\Big(\dot{x}+\phi x\Big)-\frac{96}{5}\phi x^3\Big(\alpha \dot{y}-\frac{6}{125}\phi^3\Big)-12 \alpha x^2\Big(\alpha x^3+3\dot{x}\dot{y}\Big)-48\alpha x^3 y\Big(\alpha y-\frac{8}{25}\phi^2\Big),
\end{aligned}
\end{equation*}

\begin{equation*}
\begin{aligned}
V(x,y,\dot{x},\dot{y})&=36\dot{x}^2\Big(\dot{x}+\frac{6}{5}\phi x\Big)+36 \alpha x^2 y\Big(2\dot{x}+\dot{y}\Big)+12 \alpha x^3 \Big(2\phi y -\dot{y}\Big)+\frac{144}{25}\phi^2 x^2\Big(3\dot{x}+\frac{2}{5}\phi x\Big),
\end{aligned}
\end{equation*}

\begin{equation*}\label{67b}
\begin{aligned}
I_2=&\Big(9\dot{x}^3\Big(\dot{x}+\frac{8}{5}\phi x\Big)+\frac{144}{125}\phi^3 x^3\Big(2\dot{x}+\frac{1}{5}\phi x\Big)+36 x^2 \dot{x}^2\Big(\alpha y+\frac{6}{25}\phi^2\Big)\\&+12\alpha x^3\dot{x}\Big(2 \phi y-\dot{y}\Big)-2 \alpha^2 x^4\Big(x^2+6y^2\Big)+\frac{24}{5}\phi \alpha x^4\Big(\frac{4}{5}\phi y-\dot{y}\Big)\Big) e^{\frac{12}{5} \phi t}.
\end{aligned}
\end{equation*}
One can note that we can use the method suggested in ref. \cite{VKChand} to remove the time-dependent part of the integrals, which helps to prove the integrability nature of the present cases.
\section{Invariance and Lie Symmetry analysis of the modified H\'enon-Heiles system }
The modified H\'enon-Heiles system (\ref{4a}) and (\ref{4b}) is also a Lagrangian system. The Euler-Lagrange equations of motion are
\begin{equation}\label{55a}
\frac{d}{dt}\Big(\frac{\partial L}{\partial \dot{x}}\Big)	=\frac{\partial L}{\partial x} ~\text{and}~
\frac{d}{dt}\Big(\frac{\partial L}{\partial \dot{y}}\Big)	=\frac{\partial L}{\partial y},
\end{equation}  
where $L$ is the Lagrangian given by
\begin{equation}\label{55b}
L=\Big[\frac{1}{2}\big(\dot{x}^2+\dot{y}^2\big)-\frac{1}{2}\big(Ax^2+By^2\big)-\big(\alpha x^2 y-\frac{1}{3}\beta y^3\big)\Big]e^{\phi t}.
\end{equation}
The integrability properties of the modified  H\'enon-Heiles system could also be studied through the symmetry properties of it \cite{Saha,Blu}.\\
Let $(\ref{55a})$ be invariant under the one parameter $(\epsilon)$ Lie group of infinitesimal transformations given by 

\begin{equation}\label{57a}
x \to X= x+ \epsilon~ {\eta}_1(t, x, y, \dot{x}, \dot{y}),
\end{equation}
\begin{equation}\label{57b}
y \to Y= y + \epsilon ~{\eta}_2(t, x, y,  \dot{x}, \dot{y})	,
\end{equation}
\begin{equation}\label{57c}
t \to T= t + \epsilon~ {\xi(t, x, y, \dot{x}, \dot{y})}	.
\end{equation}

This leads to the following invariance condition to be satisfied

\begin{equation}
\ddot{\eta_1}-\dot{x}\ddot{\xi}-2\dot{\xi}\Psi_1=\mathbf{X}(\Psi_1),~~\ddot{\eta_2}-\dot{y}\ddot{\xi}-2\dot{\xi}\Psi_2=\mathbf{X}(\Psi_2).
\end{equation}	

where $\Psi_1$ and $\Psi_2$ are as given in $(\ref{4a})$ and $(\ref{4b})$, respectively, and the infinitesimal operator $\mathbf{X}$ is given by
\begin{equation}
\mathbf{X}= \xi \frac{\partial}{\partial t} +\eta_1  \frac{\partial}{\partial x} +\eta_2  \frac{\partial}{\partial y} + (\dot{\eta_1}-\dot{\xi}\dot{x}) ~\frac{\partial}{\partial \dot{x}} + (\dot{\eta_2}-\dot{\xi}\dot{y}) \frac{\partial}{\partial \dot{y}} .
\end{equation}	
After a detailed calculation, we observe that\\
$\xi=0,~\eta_1=\eta_1(x, y, \dot{x}, \dot{y})e^{\lambda t},~~\eta_2=\eta_2(x, y, \dot{x}, \dot{y})e^{\lambda t},~\lambda-\text{constant},$\\
and the nontrivial symmetries occur for the P-cases only. The details of the above investigation will be published elsewhere. Using Noether's theorem, the first integral of motion can be identified for each set of Lie symmetries which takes the form
\begin{align*}
\xi=0,~\eta_1=\Big(\dot{x}+\frac{2}{5}\phi x\Big)e^{\frac{\phi}{5}t},~~\eta_2=\Big(\dot{y}+\frac{2}{5}\phi y\Big)e^{\frac{\phi}{5}t}.
\end{align*}
Similarly, the second integral of motion can be identified for each set of non-trivial symmetries of the following forms,\\
\textbf{Case 1(a):}~$A=B=\frac{6}{25}\phi^2, \beta=-\alpha.$ 
\begin{align*}
\xi=0,~\eta_1=\Big(\dot{y}+\frac{2}{5}\phi y\Big)e^{\frac{\phi}{5}t},~~\eta_2=\Big(\dot{x}+\frac{2}{5}\phi x\Big)e^{\frac{\phi}{5}t}
\end{align*}
\textbf{Case 2(b):}~$A=B=\frac{6}{25}\phi^2, \beta=-6\alpha.$ 
\begin{align*}
\xi=0,~\eta_1=\Big(4x\dot{y}-8y\dot{x}-\frac{8}{5}\phi xy\Big)e^{\frac{3}{5}\phi t},~~\eta_2=\Big(4\dot{x}x+\frac{8}{5}\phi x^2\Big)e^{\frac{3}{5}\phi t}
\end{align*}
\textbf{Case 2(c):}~$A=B=\frac{6}{25}\phi^2, \beta=-16\alpha.$ 
\begin{align*}
&\xi=0,~\eta_1=\Big(36 \dot{x}^3+\frac{216}{5}\phi x \dot{x}\Big(\dot{x}+\frac{2}{5}\phi x\Big)+\frac{288}{125}\phi^3 x^3+24\alpha x^2 y\Big(3\dot{x}+\phi x\Big)-12\alpha x^3\dot{y} \Big)e^{\frac{7}{5}\phi t},~~\eta_2=-12\alpha x^3\Big(\dot{x}+\frac{2}{5}\phi x\Big)e^{\frac{7}{5}\phi t}.
	\end{align*}

	\subsection{Separability}
	In this section, we show how the generalized symmetries obtained can be used to find coordinates transformations in which the equations of motion become separable. This is possible when the generalized symmetries are linear in velocities. 
	We say that a differentiable function $U(t,x,y,\dot{x},\dot{y})$ is an invariant for the transformations (\ref{57a}), (\ref{57b}) and (\ref{57c}) if $XU=0,$ which is a first order partial differential equation. Then its characteristic equations are 
	\begin{equation}\label{60}
		\frac{dt}{\xi}=\frac{dx}{\eta_1}=\frac{dy}{\eta_2}=\frac{d\dot{x}}
		{(\dot{\eta_1}-\dot{\xi}\dot{x})}=\frac{d\dot{y}}{(\dot{\eta_2}-\dot{\xi}\dot{y})}.
	\end{equation}
	One can also find integrals of motion to the modified H\'enon-Heiles system by solving the characteristic equation $(\ref{60})$. Since our aim is to find suitable coordinates transformations leading to separability we consider only the following part of equation $(\ref{60})$. We illustrate this for case (1)
	\begin{equation}\label{61}
		\frac{dx}{\Big(\dot{y}+\frac{2}{5}\phi y\Big)e^{\frac{\phi}{5}t}}=\frac{dy}{\Big(\dot{x}+\frac{2}{5}\phi x\Big)e^{\frac{\phi}{5}t}}.
	\end{equation}
	Solving which we get
	\begin{equation}
		(x^2-y^2)e^{-\frac{4}{5}\phi t}=constant,
	\end{equation}
	leading to the coordinates transformations
	$u=(x+y)e^{-\frac{2}{5}\phi t}, ~v=(x-y)e^{-\frac{2}{5}\phi t}.$ \\ 
	In the above coordinates the equations (\ref{1a}) and (\ref{1b}) separates into 
	\begin{equation}\label{63}
		\frac{d^2u}{dt^2}+\phi \frac{du}{dt}+A u+\alpha u^2 e^{\frac{2}{5}\phi t}=0,
	\end{equation}
	\begin{equation}
		\frac{d^2v}{dt^2}+\phi \frac{dv}{dt}+A v-\alpha v^2 e^{\frac{2}{5}\phi t}=0.
	\end{equation}
	Equation (\ref{63}) can be transformed into
	\begin{equation}\label{65}
		l^2 \frac{d^2 u}{dl^2}+2l\frac{du}{dl}+\frac{6}{25} u=-\Big(\frac{\alpha}{\phi^2}\Big) u^2 l^{\frac{2}{5}},
	\end{equation}
	by a substitution $l=e^{\phi t}$.
	Equation (\ref{65}) can be further transformed into an Emden-Fowler equation of the form
	\begin{equation}\label{66}
		\frac{d^2w}{dz^2}= - \frac{25\alpha}{\phi^2} z^{-3} w^2,
	\end{equation}
	by the substitution $l=z^5,~ u=z^{-3} w(z)$.
	The solution of equation (\ref{66}) can be written parametrically interms of the Weierstrass elliptic function \cite{And}.
	
\section{Exact solution of modified H\'enon-Heiles System}
It is well known that finding an exact solution for a given nonlinear PDE or ODE is a difficult task. Also, there exists no unique analytic method to construct an exact solution of nonlinear differential equations, in general. Recent investigations demonstrate that $\tanh$ or $\sech$  method can provide an effective tool for nonlinear ODEs and PDEs  towards constructing specific/particular solutions \cite{Wil}. In this section we extend the $\tanh$ or $\sech$ method to modified H\'enon-Heiles system. With this aim, we look for exact solution of modified H\'enon-Heiles system having the form\\
	$$ x(t)= A_1 + A_2 \tanh(bt+\delta) + A_3\tanh^2(bt+\delta),$$ $$ y(t)= B_1 + B_2 \tanh(bt+\delta) + B_3\tanh^2(bt+\delta).$$
	where $A_i,B_i, i = 1,2,3$ and $b$ are unknown constants to be determined. Substituting $x(t)$ and $y(t)$ in the equations of the modified H\'enon-Heiles system we find that each one becomes a fourth degree polynomial in $\tanh(bt+\delta)$. Then equating different powers of $\tanh(bt+\delta)$ in both the equations to zero we obtain a system of algebraic equations in $A_i$ and $B_i, i = 1,2,3$. Solving them consistently yields the following exact solutions. 
	
	Exact solution for integrable cases:
	\begin{equation*}
		\begin{aligned}
			\textbf{(i)}~A&=B= \frac{6}{25} \phi^2 ,~ \beta = -\alpha,\\
			x(t)&=\frac{3\phi^2}{100\alpha}\Big[3+\tanh{\left(-\frac{\phi}{10}t+\delta\right)\Big]}\Big[-1+\tanh\left(-\frac{\phi}{10}t+\delta\right)\Big],\\
			y(t)&=-\frac{3\phi^2}{100\alpha}\Big[\tanh^2\left(-\frac{\phi}{10}t+\delta\right)+2\tanh\left(-\frac{\phi}{10}t+\delta\right)+5\Big].\\
			\textbf{(ii)} ~A&= B =\frac{6}{25}\phi^2, ~ \beta = -6 \alpha,\\			
			x(t)&= \frac{3I\phi^2}{25 \alpha}\Big[(1+\tanh\left(-\frac{3\phi}{10}t+\delta\right)\Big],\\
			y(t)&=\frac{-3\phi^2}{100 \alpha}\Big[1+\tanh\left(-\frac{3\phi}{10}t+\delta\right)\Big] \Big[3\tanh\left(-\frac{3\phi}{10}t+\delta\right)-1\Big].\\
   \textbf{(iii)}~ A&=\dfrac{21}{100}\phi^2, B=-\dfrac{24}{100}\phi^2,~  \beta=-16\alpha,~\alpha=-\frac{3}{50}\phi^2,\\
			x(t)&=-\frac{I \sqrt{14}}{2}\Big[\tanh\left(-\frac{\phi}{10}t+\delta \right)+1\Big]^2,\\
			y(t)&=\frac{1}{4}+\tanh\left(-\frac{\phi}{10}t+\delta \right)+\frac{1}{2}\tanh^2\left(-\frac{\phi}{10}t+\delta\right).
		\end{aligned}
	\end{equation*}
	Exact solution for other cases:
	\begin{equation*}
		\begin{aligned}		
			\textbf{(iv)} ~A&= B =-\frac{4}{9} \phi^2,~ \beta = -\frac{4}{3}\alpha,\\
			x(t)&=\frac{\phi^2}{6\sqrt6\alpha}\Big[\tanh\left(\frac{\phi}{6}t+\delta\right)+1\Big]\Big[\tanh\left(\frac{\phi}{6}t+\delta\right)-1\Big],
			\\ y(t)&=\frac{-\phi^2}{12\alpha}\Big[\tanh\left(\frac{\phi}{6}t+\delta\right)+1\Big]\Big[\tanh\left(\frac{\phi}{6}t+\delta\right)-3\Big].\\	
			\textbf{(v)} ~A&=\frac{9}{100} \phi^2, B=-\frac{6}{25} \phi^2,~ \beta = -\frac{16}{5}\alpha,\\
			x(t)&= \frac{3I\sqrt30 \phi^2}{500\alpha}\Big[\tanh\left(-\frac{\phi}{10}t+\delta\right)+1\Big]^2,\\
			y(t)&=-\frac{3\phi^2}{200\alpha}\Big[-3 + 4 \tanh\left(-\frac{\phi}{10}t+\delta\right) + 2\tanh^2\left(-\frac{\phi}{10}t+\delta\right)\Big].
		\end{aligned}
	\end{equation*}
	We see that cases $\textbf{(i)}$ and $\textbf{(ii)}$ above have an exact solution for the integrable cases $~(1a)$ and $~(2b)$ respectively. Though the integrability of parametric cases $~(1b)$ and $~(2a)$ is not established, an exact solution for the same is given in
	$\textbf{(iv)}$ and $\textbf{(v)}$ respectively, which may correspond to the cases admitting lesser number of arbitrary constants in the Laurent series. It appears that  case$~1(c)$  may not admit the exact solution expressed in $\tanh$ or $\sec$h function. One may also be  able to find particular forms of elliptic functions, generalizing the above procedure which we hope to explore in future.
	
\section{Summary and concluding remarks}
	In this article, by applying the Painlev\'e analysis of ordinary differential equations to modified H\'enon-Heiles system we report that it admits the Painlev\'e property for three distinct parametric restrictions. For each of the identified cases, we construct two independent integrals of motion using the well known Prelle-Singer method. Thus we infer that the modified H\'enon-Heiles system is integrable for three distinct parametric restrictions. We have shown that there exists  an exact solution  for two of the isolated three integrable cases of modified H\'enon-Heiles system expressed in terms of $\tanh$ function.
	We observe that when
	\begin{equation}
		\beta=-16\alpha,\  \phi_2=\phi_1=\phi \label{*}
	\end{equation}
 
leads to the integrable case $~2(c)$, where \\
	\hspace*{5cm} $A=B=\dfrac{6}{25}\phi^2.$ \\
	But we are able to find an exact solution expressed in $\tanh$ function for the case in \eqref{*}  when
	$A=\dfrac{21}{100}\phi^2,\ B=-\dfrac{24}{100}\phi^2,~ \alpha=-\dfrac{3}{50}\phi^2$.
	\indent
	Also, we have isolated two more parametric restrictions of modified H\'enon-Heiles system admitting Laurent series solution with two arbitrary constants and two of them admit an exact solution expressed in $\tanh$ function. From the investigation, we observe that the inclusion of damping term in the equations of motion leads to further parametric restriction in the linear terms of the equations. The connection between the Laurent series solution and the $\tanh$ solution of the integrable case needs further investigation.\\
	
\noindent\textbf{Acknowledgments}\\
The authors thank the anonymous referees for their constructive comments. N. Muthuchamy would like to thank CSIR, New Delhi, India for providing financial support in the form of CSIR-SRF. The work of V. K. Chandrasekar forms part of a research project sponsored by SERB DST-CRG Project Grant No. C.R.G./2020/004353. V. K. Chandrasekar also thanks DST, New Delhi, for computational facilities under the DST-FIST programme (Grant No. SR/FST/PS-1/2020/135) to the Department of Physics. The work of M. Lakshmanan is supported by the DST-SERB National Science Chair position.\\
	
	\section*{AUTHOR DECLARATIONS}
	
\noindent \textbf{Author Contributions}\\
\textbf{C. Uma Maheswari}: Methodology, Supervision, Writing - Review and Editing, \textbf{N. Muthuchamy}: Writing – Methodology-Original Draft.\textbf{ V. K. Chandrasekar}: Writing - Review and Editing, Supervision.  \textbf{ R. Sahadevan}     : Conceptualization, Methodology. \textbf{M. Lakshmanan:} Conceptualization, Methodology, Writing - Review and Editing.\\

\noindent \textbf{Data Availability}\\	Data sharing is not applicable to this article as no new data were created or analyzed in this study.\\

\noindent \noindent\textbf{Conflict of Interest} \\The authors declare that they have no conflict of interest.

\nocite{*}
\section*{References}
\bibliography{HenonHeiles.bib}

\end{document}